\documentclass[final,3p,times,twocolumn]{elsarticle}
\usepackage{lineno}
\linenumbers

\usepackage{epsfig}
\usepackage{epstopdf}
\usepackage{rotating}
\usepackage{amssymb}

\journal{Nuclear Instruments and Methods in Physics Research A}

\begin{document}

\begin{frontmatter}

\title{Measurement of the dependence of the light yields of linear alkylbenzene-based and EJ-301 scintillators on electron energy}

\author{\corref{cor1}H.~Wan Chan Tseung}
\author{J.~Kaspar}
\author{N.~Tolich}
\cortext[cor1]{Corresponding author. Tel: +1 206 543 4035; fax: +1 206 685 4634. {\it Email address}: hwan@uw.edu}

\address{Center for Experimental Nuclear Physics and Astrophysics, and Department of Physics, University of Washington, Seattle, WA 98195}

\begin{abstract}
An experimental test of the electron energy scale linearities of SNO\protect\raisebox{.15ex}{+} and EJ-301 scintillators was carried out using a Compton spectrometer with electrons in the energy range 0.09--3 MeV. The linearity of the apparatus was explicitly demonstrated. It was found that the response of both types of scintillators with respect to electrons becomes non-linear below $\sim$0.4 MeV.  An explanation is given in terms of Cherenkov light absorption and re-emission by the scintillators. 


\end{abstract}

\begin{keyword}
scintillator, linearity, energy scale, linear alkylbenzene, SNO\protect\raisebox{.15ex}{+}


\end{keyword}

\end{frontmatter}


\section{Introduction}
\label{}

SNO+ is a multi-purpose neutrino experiment whose reach extends to the following areas of neutrino physics: neutrinoless double beta decay (with Nd-loaded scintillator), geo-neutrinos, reactor and low-energy solar neutrinos, as well as supernova neutrinos \cite{Chen}. It is a $\sim$780-tonne liquid scintillator (LS) detector currently under construction $\sim$2 km underground at the SNOLAB facility in Sudbury, Ontario, Canada. The scintillator will be contained in a 12-m diameter spherical acrylic vessel surrounded by $\sim$9500 photomultiplier tubes (PMTs). More details on the layout can be found in \cite{SNO}.

The LS to be used in SNO\protect\raisebox{.15ex}{+} is linear alkylbenzene (LAB) with $\sim$2 g/L of PPO (2,5-diphenyloxazole). In the double-beta decay phase, the LAB-PPO will be loaded with $\sim$0.1 \% Nd (by weight). To achieve the goals of the experiment, it is imperative to understand the properties of this scintillator down to the SNO\protect\raisebox{.15ex}{+} energy threshold, which is expected to be $\sim$0.2 MeV for low-energy solar neutrino studies. In particular, for electrons, a precise knowledge of the dependence of the light yield,  ${\cal{L}}$,  on their kinetic energy, $E_e$, is required. For scintillators, ${\cal{L}}(E_e)$ is commonly assumed to be linear, except at very low energies (typically below $\sim$50 keV), where the effects of ionization quenching appear to be important. The decrease in light yield due to ionization quenching is well-described for all particles by Birks' law \cite{Birks}. 

In the last decade, the KamLAND experiment observed, using $\gamma$-ray calibration sources, that the electron energy scale of a dodecane-based scintillator was non-linear and could not be solely described by Birks' law \cite{kamlandthesis}. 
A laboratory investigation of the KamLAND LS electron energy scale with a Compton spectrometer confirmed these results \cite{Oleg}. It was hypothesized that Cherenkov light, being absorbed and re-emitted by the LS, acted as another scintillation light `source' besides ionization. This led to a rise in $\mathrm{d}{\cal L}/\mathrm{d}E_e$ in the energy region above the Cherenkov threshold for electrons. 
 
The primary aim of this work is to test the linearity of ${\cal{L}}(E_e)$ for both LAB-PPO and Nd-doped LAB-PPO, in an energy range relevant to the various goals of the SNO\protect\raisebox{.15ex}{+} experiment, {\it i.e.} up to 3.5 MeV. We verify if the SNO\protect\raisebox{.15ex}{+} LS displays the same behaviour as KamLAND's, and quantify  the extent of the Cherenkov contribution to non-linearity. We also test the linearity of EJ-301 scintillator \cite{Eljen}, which is equivalent to the commonly-used NE-213. 
Details of our apparatus and method are given in  \S\ref{section:experimentalmethod}. \S\ref{section:linearity} describes a measurement of the non-linearity of our set-up. The results of our scintillator linearity tests are presented and discussed in \S\ref{section:results}. We summarize and conclude in \S\ref{section:conclusion}.

 \section{Experimental method}\label{section:experimentalmethod}
\begin{figure}[!pt]\centering
\includegraphics[width=6.3cm]{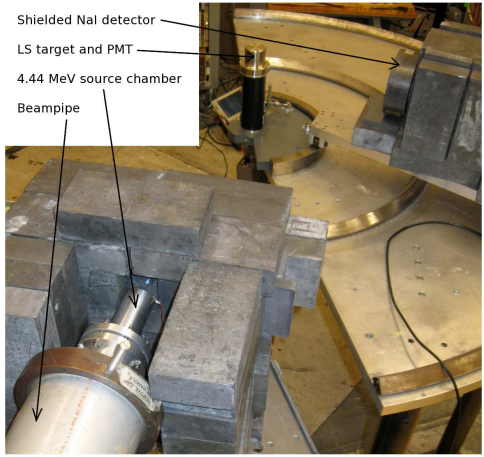}
\caption[]
 {Set-up of the apparatus. }
 \label{figure:compton_apparatus}
 \end{figure}

To measure the scintillation light output as a function of $E_e$, we used the Compton scattering of $\gamma$-rays of known energies, $E_{\gamma}$, to produce mono-energetic electrons within the LS. $E_e$ is given by:
 \begin{equation}E_e=E_{\gamma}-E_{\gamma}^{\prime}\end{equation}
 where $E_{\gamma}^{\prime}$ is the scattered $\gamma$ energy. $E_{\gamma}^{\prime}$ is  calculated using the Compton formula:
\begin{equation}\label{equation:compton} E_{\gamma}^{\prime}=\frac{E_{\gamma}}{1+\frac{E_{\gamma}}{m_e c^2}(1-\cos\theta)}\end{equation}
where $m_e$ is the electron mass, and $\theta$ is the $\gamma$ scattering angle (defined in the lower part of fig.~\ref{figure:compton_apparatus}). The scattered $\gamma$ was recorded with a NaI detector placed at angle $\theta$. Only events that resulted in coincident pulses from the LS and NaI detectors were considered. 
Two $\gamma$-ray sources of 0.662 and 4.44 MeV allowed values of $E_e$ in the range 0.09--3 MeV to be probed with high granularity. 

Fig.~\ref{figure:compton_apparatus} is a photograph of the spectrometer set-up, with the LS cell located 1.5 m from the $\gamma$ source, and a $2''$ NaI detector placed on a trolley with a 0.5-m rotation radius. The $\gamma$-rays were collimated by a 5-mm diameter hole drilled through 20 cm of lead shielding. 
The 0.662 MeV $\gamma$s originated from a $^{137}$Cs source that was small enough to fit into the collimator hole. The 4.44 MeV source is described below. 

\subsection{A 4.44 MeV $\gamma$-ray source}
We constructed a high intensity 4.44 MeV  $\gamma$ source using the $^{12}$C(p,p$'$)$^{12}$C reaction. A $\sim$2 $\mu$A 5.7 MeV proton beam was supplied by the CENPA tandem Van de Graaff accelerator. Our source chamber, seen in fig.~\ref{figure:compton_apparatus}, was made of aluminium. It consisted of a flange reducer section that interfaced to the beam pipe, and an insulated target holder section connected to a pico-ammeter for flux monitoring. A small tantalum aperture located at the beam pipe interface ensured that most $\gamma$s were produced at the centre of the carbon target, which was a 5-mm thick disk of natural carbon. Cooling for the latter was not required. We carefully aligned the LS  cell, $\gamma$-ray collimator, and the proton beam collimating and focussing apertures with a surveying telescope located far away in the $0^{\circ}$ direction.

The $^{12}$C(p,p$'$)$^{12}$C cross-section was previously measured by Barnard {\it et al.} and Dyer {\it et al.} \cite{Barnard, Dyer}. The beam was operated at 5.7 MeV to select the first resonance peak and reduce backscattered $\gamma$ backgrounds in the experimental hall as much as possible. At 2 $\mu$A of 5.7 MeV proton beam current, and assuming a cross-section of 200 mb, we estimated the overall source strength to be a few tens of mCi.

A NaI detector placed along the beamline and 2 m from the Carbon target, confirmed the intense 4.44 MeV gamma event rate. 
Two other photons, with energies of 3.11 and 0.511 MeV, were emitted from $^{13}$C(p,p$'$)$^{13}$C, and from the decay of $^{13}$N produced in the $^{13}$C(p,n)$^{13}$N reaction. The relative contributions of the 4.44, 3.11 and 0.511 MeV $\gamma$s are 89.1, 0.9 and 10\%, respectively \cite{Guldbakke}. At any angle $\theta$, the 
values of $E_e$ for recoil electrons in the LS from the 0.511 MeV and 4.44 MeV $\gamma$s were very different, while the 3.11 MeV $\gamma$ interfered with the electron signal from the 4.44 MeV line at the $<1$\% level. Therefore, the 0.511 and 3.11 MeV lines can be neglected in this work.
 
 \subsection{Scintillator cell}
Two scintillator-PMT assemblies were custom-designed and built with all materials in direct contact with the active volume being compatible with the LAB-based and EJ-301 scintillators. The scintillator was coupled to a Philips Valvo XP-2262 PMT, housed in mu-metal shielding, via a pyrex window. The PMT was biased to $-1400$ V. The $2''\times2''$ cylindrical scintillator container was bored out of a solid piece of stainless steel. Its walls were 1 mm thick, with mirror-polished inner surfaces. 

Both of our LAB-based LS samples contained 3 g/L of PPO. Before filling, the LS was de-oxygenated by gently bubbling in Ar gas for $\sim$45 minutes \cite{Xiao}. The cells were  cleaned in an ultra-sound bath, before being vacuum-filled with the LS. We used tubes and valves that were made entirely of teflon. 

\begin{figure}[!pt]\centering
\includegraphics[width=8cm]{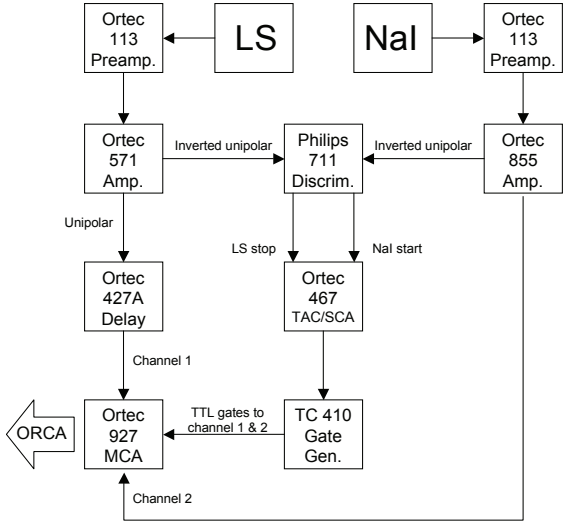}
\caption[]
 {Schematic of the electronics chain. }
 \label{figure:electronics_flowchart}
 \end{figure}

\subsection{Electronics and data acquisition}

An NIM-based electronics set-up (shown in fig.~\ref{figure:electronics_flowchart}) recorded the charge of coincident PMT pulses from the NaI and liquid scintillator detectors. The pulse height spectra from the multi-channel analyzer (MCA) were read in by ORCA, a data-acquisition software developed at the Universities of Washington and North Carolina \cite{ORCA}. 

\subsection{Data-taking procedure}
The electronics and PMT were allowed to stabilize overnight before data-taking the next day. To avoid making corrections due to gain drifts, measurements with the 4.44 MeV and 0.662 MeV $\gamma$s were done on the same day, with the $^{137}$Cs runs starting soon after switching off the proton beam.  For consistency, the LAB-PPO and Nd-doped LAB-PPO measurements were performed using the same scintillator cell and the same PMT, which underwent the linearity tests described in \S\ref{section:linearity}. The electronics settings were fixed throughout the whole experiment.

\begin{figure}[!pt]\centering
\includegraphics[width=8cm]{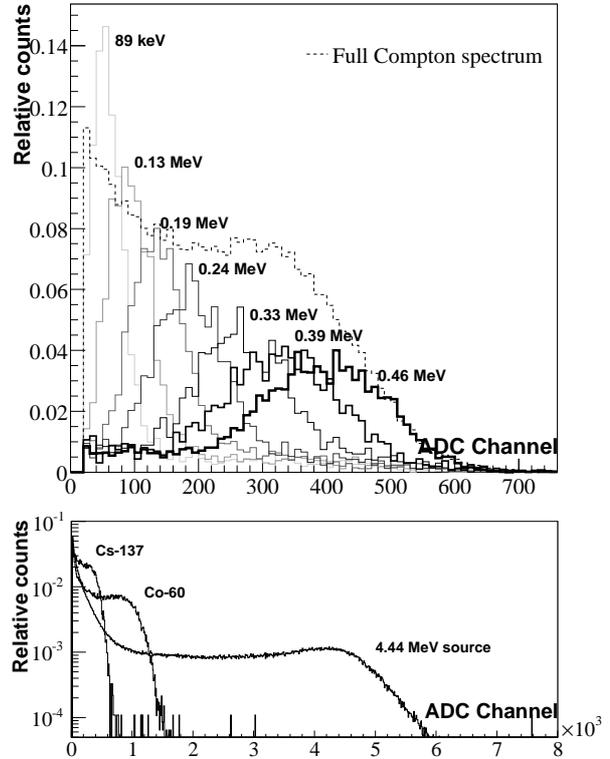}
\caption[]
 {Top: From left to right, the solid-line histograms are the observed electron recoils in LAB-PPO resulting from the 28.5$^{\circ}$, 36$^{\circ}$, 46$^{\circ}$, 56$^{\circ}$, 76$^{\circ}$, 96$^{\circ}$ and 141$^{\circ}$ scattering of $\gamma$-rays originating from a $^{137}$Cs source. The reconstructed values of $E_e$ at each angle are given. The dotted line is the full Compton scattering spectrum. Bottom: The LAB-PPO Compton spectra of 0.662, $\sim$1.25 and 4.44 MeV $\gamma$s from $^{137}$Cs, $^{60}$Co and the $^{12}$C(p,p$'$)$^{12}$C reaction, respectively. }
 \label{figure:exampleData}
 \end{figure}

Fig.~\ref{figure:exampleData} shows the LS pulse height spectra at seven angles between $28.5^{\circ}$ and $141^{\circ}$. This data was taken with the $^{137}$Cs source and LAB-PPO target. At each angle, mono-energetic electron peaks can unambiguously be located to give ${\cal L}$ in units of ADC bins. The full Compton spectrum, drawn in dotted line, is shown for comparison. The bottom plot shows two more Compton spectra,  obtained with a  $^{60}$Co source and our 4.44 MeV accelerator source.

\section{Linearity of apparatus}\label{section:linearity}
The recorded ADC peak positions from the LS can scale non-linearly with $E_e$ for the following reasons: (1) the scintillator response is intrinsically non-linear, (2) the integrated charge from the XP-2262 PMT pulses did not increase linearly with the number of photons incident on its photocathode, (3) the electronics chain distorted the energy scale, and (4) the apparatus was misaligned, resulting in erroneous values of $E_e$ when using the Compton equation with scattering angles that were not corrected for this misalignment. The last three issues are discussed below.  

\subsection{Misalignment and electron energy reconstruction}\label{section:energyreconstruction}
\begin{figure}[!pt]\centering
\includegraphics[width=8cm]{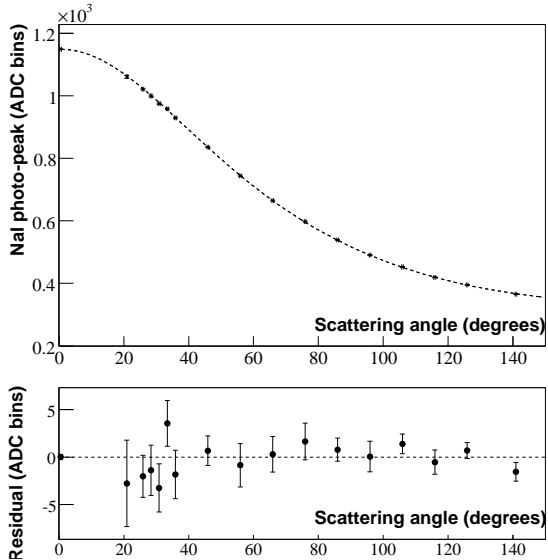}
\caption[]
 {Top: The angular dependence of NaI photo-peak positions (black markers) of $\gamma$s from a $^{137}$Cs source, scattered off a Nd-doped LAB-PPO target. The dotted line is a fit to Eq.~\ref{equation:comptonfit}. Bottom: Fit residuals.}
 \label{figure:AllNaIdata}
 \end{figure}

The spectrometer misalignment can be studied by fitting the angular dependence of NaI photo-peak positions, $f(E_{\gamma}^{\prime})$, from $^{137}$Cs source data to the Compton formula, Eq.~\ref{equation:compton}. The fit function was:
\begin{equation}\label{equation:comptonfit}f(E_{\gamma}^{\prime})=\frac{a\cdot E_{\gamma}}{1+\frac{E_{\gamma}}{m_e c^2}(1-\cos(\theta+\delta))}+b\end{equation}
where $a$ and $b$ are constants, and $\delta$ is an angular offset to account for the misalignment.
 Fig.~\ref{figure:AllNaIdata} shows NaI data (black markers) fitted to Eq.~\ref{equation:comptonfit} from $0^{\circ}$ to $141^{\circ}$. The uncertainty on reading out $\theta$ was estimated to be $\pm 0.25^{\circ}$. Except at $\theta=0^{\circ}$, all the data were collected in coincidence with the LS. 
When $\delta$ was fixed to $0^{\circ}$, the $\chi^2$/NDF for the fit was 19.9/15. When it was allowed to float, the best fit value for $\delta$ was $-0.36^{\circ}\pm 0.13^{\circ}$, with an improved $\chi^2$/NDF of 7.8/14. To account for this angular offset, we increase our uncertainty in the scattering angle from $\pm 0.25^{\circ}$ to $\pm 0.36^{\circ}$. The residuals for the unconstrained fit are shown in the lower plot of fig.~\ref{figure:AllNaIdata}. 

We performed further checks on the reconstructed $E_e$ values using another method, which consists of calibrating the NaI energy scale with $^{137}$Cs, $^{207}$Bi and $^{152}$Eu $\gamma$-ray sources. A fit to a straight line below $0.7$ MeV yielded values of $a$ and $b$ that were consistent with the Compton fit. The calibration constants were then used to get $E_{\gamma}^{\prime}$, and hence $E_e$. This method is independent of the one described in \S\ref{section:experimentalmethod}, because no assumptions are made about Compton scattering. The mean percentage difference between the values of $E_e$ reconstructed using the two methods was $-0.4$ \% for $^{137}$Cs data with a LAB-PPO target, and $1$ \% for a Nd-doped LAB-PPO target. 

Note that the second method is applicable only when the NaI photo-peaks for the scattered $\gamma$s are clearly visible. For the 4.44 MeV $\gamma$ data, this was not the case, and it was assumed Eq.~\ref{equation:compton} still accurately applied.

\subsection{Measurement of the linearity of PMT and electronics chain}\label{section:pmtlinearity}
\begin{figure}[!pt]\centering
\includegraphics[width=8cm]{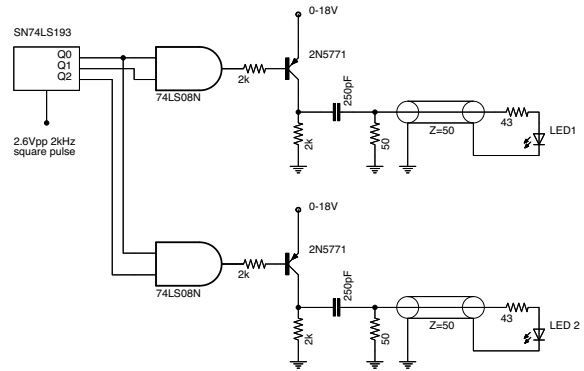}
\caption[]
 {Schematic of the LED driver circuit. }
 \label{figure:led_driver_schematic}
 \end{figure}
 
For the LAB-based LS, the linearity of the PMT and electronics chain was verified using a system of two sequentially flashing Light Emitting Diodes (LEDs), following the method described in \cite{Tripathi}. The two LEDs (diffuse yellow HLMP-3401) were placed adjacent to each other, 10 cm in front of the XP-2662 in a light-tight box. The PMT was connected to exactly the same electronics (with the same voltage and gain settings) as in the scintillator linearity measurements. An LED driver (fig.~\ref{figure:led_driver_schematic}) continuously flashed the LEDs in the following sequence consisting of three steps: (A) LED 1 only, (B) LED 2 only, and (C) LEDs 1 and 2 at the same time. The driver circuit consisted of a 4-bit binary counter driven by an Agilent 33120a pulser. The Q$_0$, Q$_1$ and Q$_2$ outputs from the counter were fed to two AND gates as shown in fig.~\ref{figure:led_driver_schematic}. The logic outputs operated PNP transistor switches that were  capacitively coupled to LEDs 1 and 2. The resulting pulses were 20 ns wide at full-width-half-maximum, with a variable amplitude of 0--4 V. The pulse amplitudes, and thus the number of photons incident on the PMT, were  controlled by an Instek GPS-3303 DC power supply, which biased the emitters of the two transistor switches. To reduce reflections, a 43 $\Omega$ resistor was placed in series with the LEDs, which had a forward bias resistance of $\sim$6 $\Omega$.

\begin{figure}[!pt]\centering
\includegraphics[width=8cm]{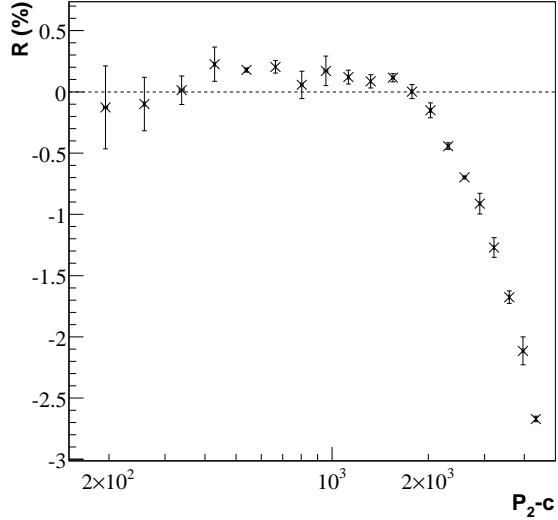}
\caption[]
 {Measured non-linearity of the PMT and electronics chain. The PMT that underwent the test was the one used in the LAB-based scintillator measurements.}
 \label{figure:linearityplot}
 \end{figure}

The voltage pulses during steps A and C, as well as B and C were observed to be nearly identical. The following ratio $R$ quantifies non-linearity:
\begin{equation}\label{equation:nonlinearR} R=100\cdot\frac{(P_3-P_1)-P_2+c}{P_2-c}\end{equation}
where $P_1$, $P_2$ and $P_3$ are the ADC peak positions in steps A, B and C respectively, and $c$ is the pedestal in number of ADC channels. The power supply voltages to the two transistors were varied such that $P_1\!\!<\!\!P_2$ and $P_1/P_3\approx 1/3$ and $P_2/P_3\approx 2/3$. Fig.~\ref{figure:linearityplot} shows $R$ as a function of $P_2-c$. The difference between  $P_3$ and $P_1$ is consistent with $P_2$ at the 0.2\% level,  except at high ADC channels, when space charge effects in the PMT become important. The resulting gain loss makes $P_3$ lower than $P_1+P_2$. We will discuss our results in terms of the non-linearity parameter $R$ in the next section. 

$c$ was measured by injecting an exponentially-decaying pulse of controllable amplitude into the Ortec 571 amplifier (fig.~\ref{figure:electronics_flowchart}). It was found to be $-10.72\pm 0.73$. In the process of measuring $c$, we confirmed the linearities of the Ortec 571 amplifier, 427A delay and 927 MCA units that were used to record pulse heights from the LS PMT. 

\section{Results}\label{section:results}

\begin{sidewaysfigure*}[p]
\centering
\includegraphics[width=7.2cm]{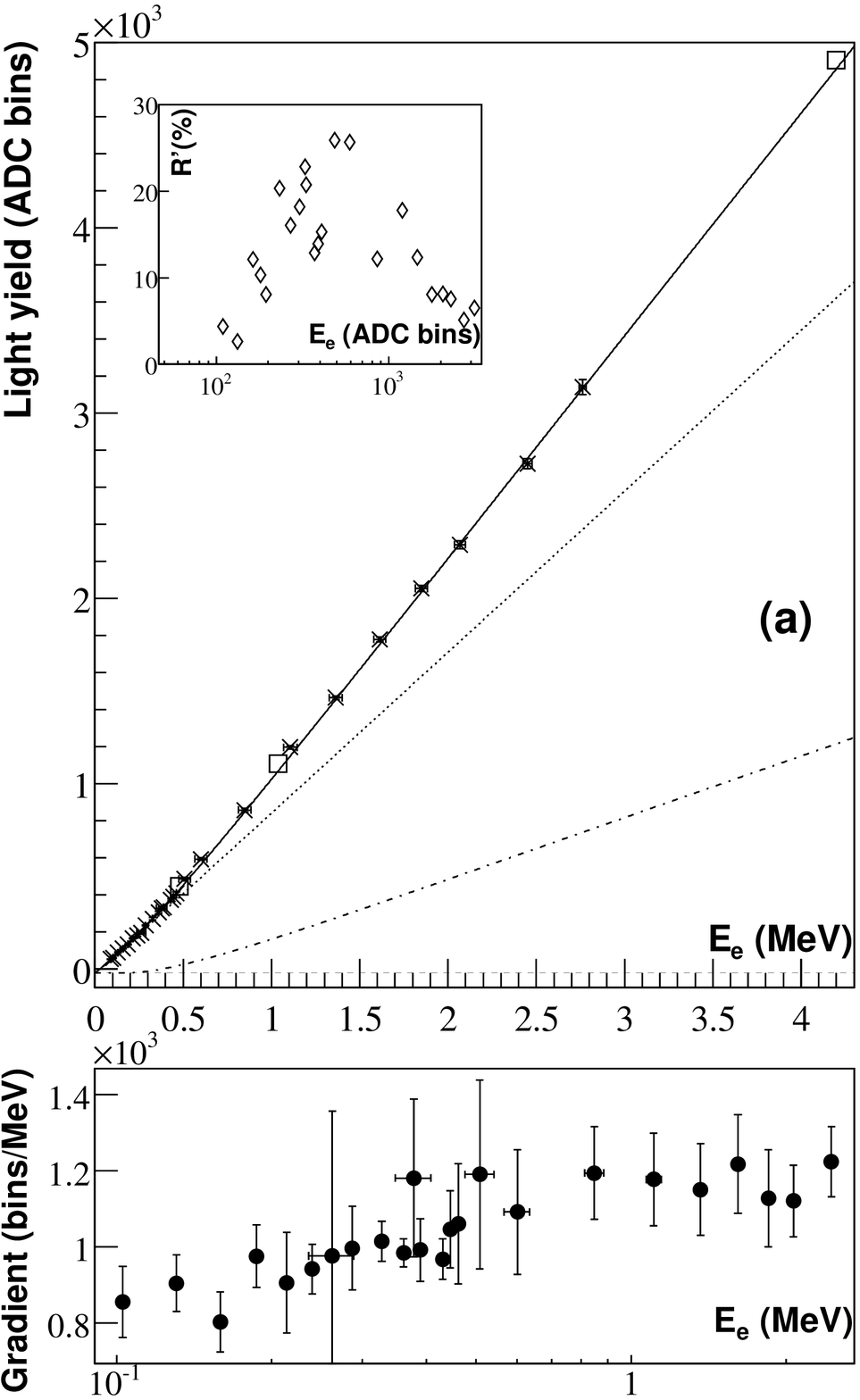}
\includegraphics[width=7.2cm]{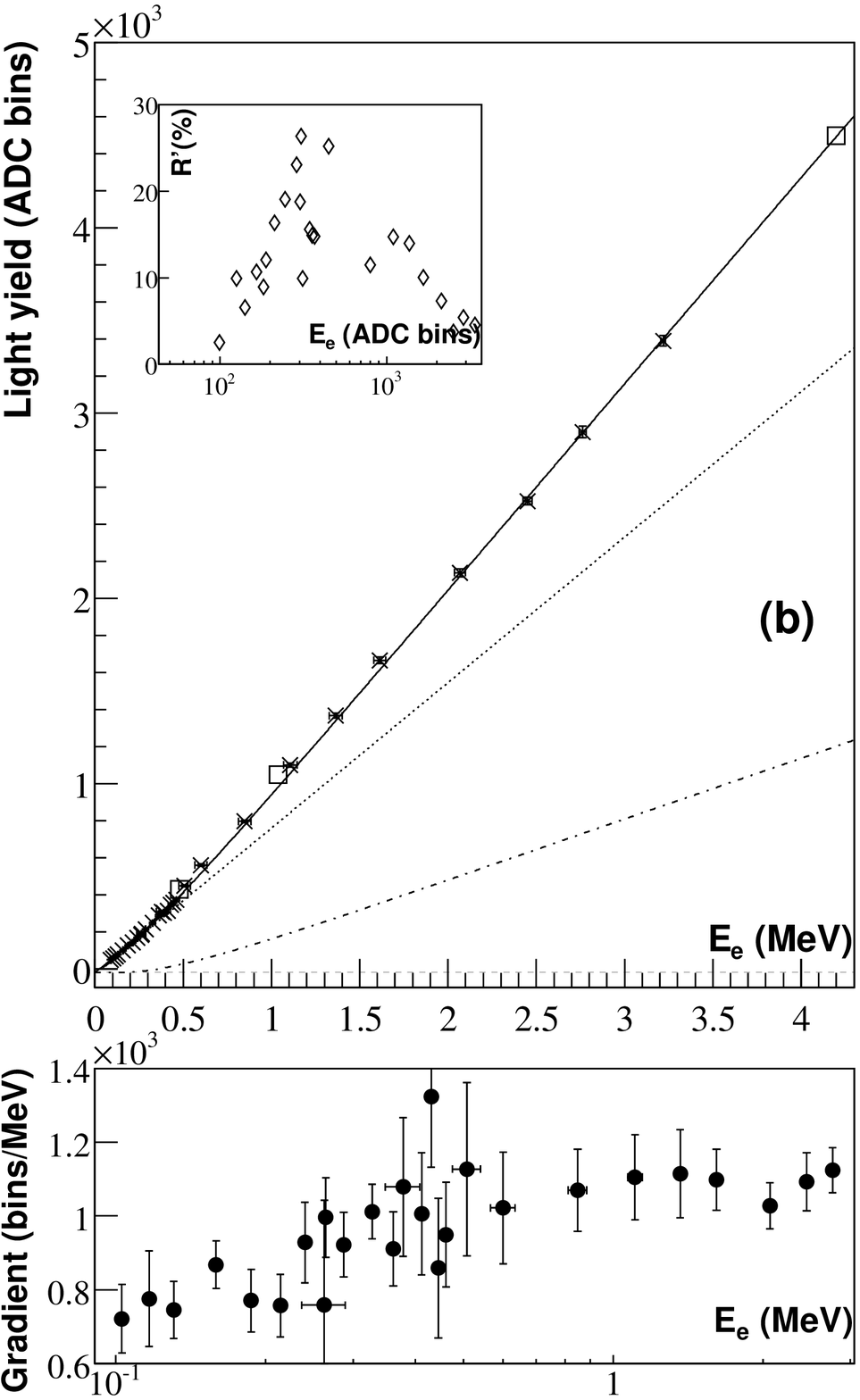}
\includegraphics[width=7.2cm]{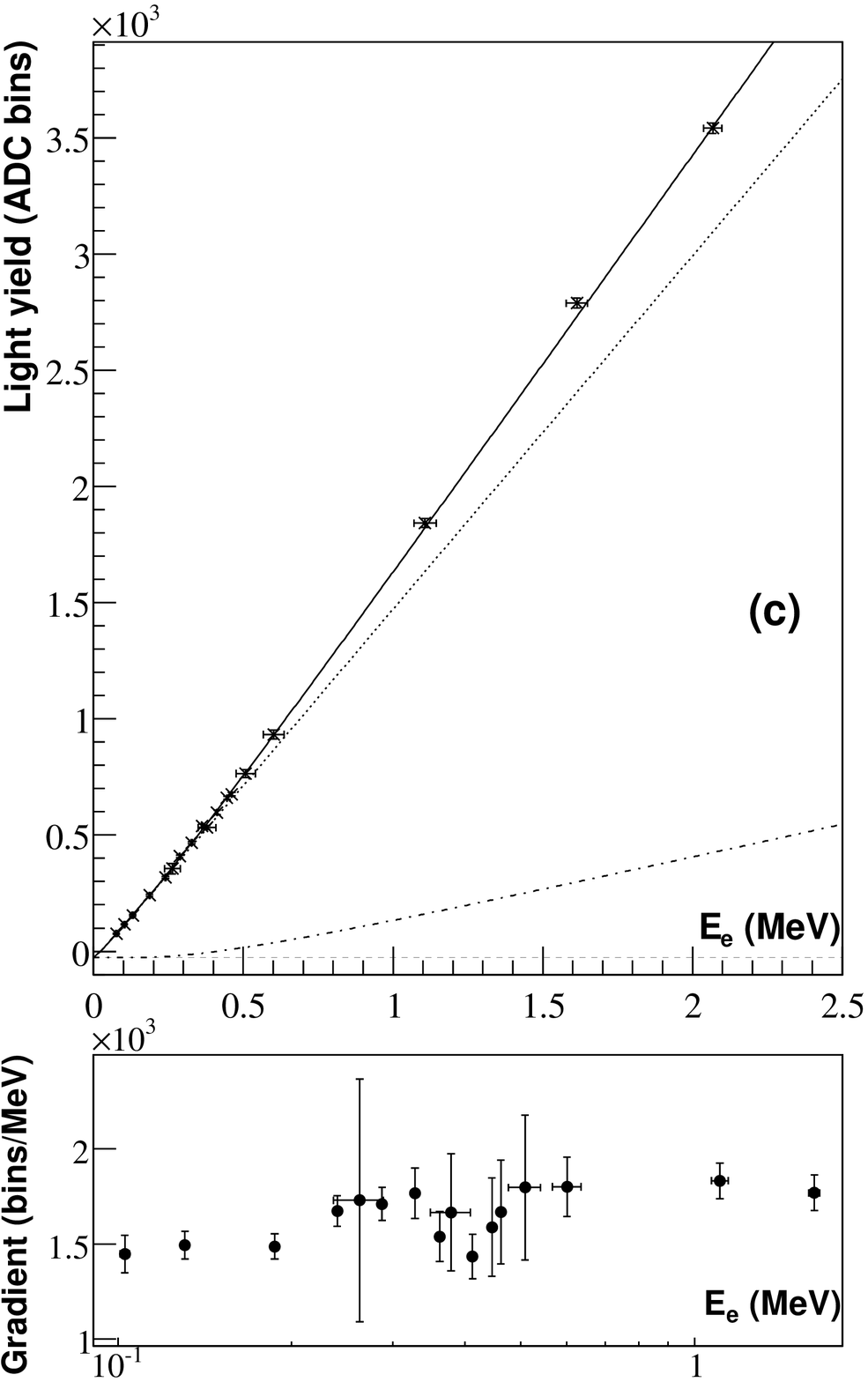}
\caption[]
 {(a) The LAB-PPO, (b) Nd-doped LAB-PPO and (c) EJ-301 electron energy scales ${\cal L}(E_e)$ in units of ADC bins. The Compton spectrometer measurements are displayed with the cross markers in the upper plot. The circle markers in (a) and (b) are the Compton edge positions of $\gamma$-rays from $^{137}$Cs, $^{60}$Co, and $^{12}$C(p,p$'$)$^{12}$C. The solid-line curves are the best fits to a simple light yield model (Eq.~\ref{equation:fitequation}). The scintillation and Cherenkov components are shown as the dotted and dash-dotted curves, respectively. The lower plots show the change in $\mathrm{d}{\cal L}/\mathrm{d}E_e$  with $E_e$, while the insets show $R'$ (defined in Eq.~\ref{equation:rprime}) as a function of $E_e$.}
 \label{figure:resultsdatafit}
 \end{sidewaysfigure*}

Our measurements of ${\cal L}(E_e)$ in ADC counts, for LAB-PPO (26 data points), Nd-doped LAB-PPO (29 data points) and EJ-301 (18 data points), are shown in cross markers in figs.~\ref{figure:resultsdatafit}(a), (b) and (c), respectively. For the LAB-based scintillators, $\mathrm{d}{\cal L}/\mathrm{d}E_e$, shown in the lower plots, increases by nearly 50 \% in the energy range from 0.2 to 1 MeV. The change in $\mathrm{d}{\cal L}/\mathrm{d}E_e$ is less pronounced for EJ-301. In figs.~\ref{figure:resultsdatafit}(a) and (b), the circle markers are the 0.48, 1.04 and 4.2 MeV Compton edge positions (evaluated at half-height) for $\gamma$s from $^{137}$Cs, $^{60}$Co, and $^{12}$C(p,p$'$)$^{12}$C, respectively. These three data points serve as a further verification of our alignment and energy reconstruction procedure.

In \S\ref{section:pmtlinearity}, we discussed a measurement of the non-linearity $R$ (Eq.~\ref{equation:nonlinearR}) due to the PMT and electronics at 15 $P_2$ values between 200 and 3500 ADC bins. This involved using two light sources (LEDs), one of which was nearly twice as strong as the other. In the LS, the electron is also a light source. For $200\!\!<\!\!{\cal L}(E_e)\!\!<\!\!3500$, we calculated the analogous quantity $R^{\prime}$ by interpolation: 
\begin{equation}\label{equation:rprime}  R^{\prime}=100\cdot\frac{\left[{\cal L}(E_e)-{\cal L}(\frac{E_e}{3})\right]-{\cal L}(\frac{2E_e}{3})+c}{{\cal L}(\frac{2E_e}{3})-c}\end{equation}
If the LS is linear with respect to $E_e$, $R^{\prime}$ should be of the same order as $R$. The insets in figs.~\ref{figure:resultsdatafit}(a) and (b) show the values of $R^{\prime}$ as a function of $E_e$. For both LAB LS  samples, these clearly exceed 0.2\%, reaching $\sim$20 \% in the low-energy region. Therefore, the non-linearities seen in figs.~\ref{figure:resultsdatafit}(a) and (b) were not caused by the PMT and electronics. We assume this to be also the case for  fig.~\ref{figure:resultsdatafit}(c), although $R$ has not been measured for the PMT that was used in the EJ-301 measurements.

\subsection{Interpretation and discussion}
We now provide a simple model for ${\cal L}(E_e)$. At any value of $E_e$, the total light yield of the LS, in a wavelength region detectable by the PMT, is given by:
\begin{equation}\label{equation:lightyield}N_T=N_S+N_C\end{equation}
where $N_S$ is the number of photons originating from the scintillation process and $N_C$ is the Cherenkov contribution. $N_C$ includes primary Cherenkov photons, as well as those emitted as a result of LS excitation by Ultra-Violet (UV) Cherenkov photons. ${\cal L}(E_e)$ is proportional to $N_T$. We fitted our data to the following:
\begin{equation}\label{equation:fitequation} {\cal L}(E_e)=A\cdot N_S^{\prime}(E_e)+B\cdot N_{C2}(E_e)+C\end{equation} 
where $A$ and $B$ are scaling constants, and $C$ is an offset. These take into account detector-related effects such as photon collection and PMT quantum efficiencies, gain, and pedestal. The scintillation component is given by:
\begin{equation} A\cdot N_S^{\prime}=A\cdot\int_{E_e}^{0}\frac{1}{1+k_B S_e(E_e^{\prime})}\mathrm{d}E_e^{\prime}\end{equation}
where $k_B$ is Birks' constant, and $S_e$ is the electron stopping power, calculated using the ESTAR database \cite{ESTAR}. Since the scintillation light yield is unknown, it was factored into $A$. $N_{C2}(E_e)$ was calculated with:
\begin{equation}\label{equation:cerenkov} N_{C2}=\int_{\omega_1}^{\omega_2}\!\!\!\!\int_{E_e}^{E_f}\!\!\!\frac{\alpha}{S_e(E_e^{\prime})}\!\left(1-\frac{1}{\beta^2(E_e^{\prime})n^2(\omega)}\right) \mathrm{d}E_e^{
\prime}\mathrm{d}\omega\end{equation}
where $\beta$ is the electron speed in units of $c$, and $E_f$ is defined by $\beta(E_f)=1/n(\omega)$. In Eq.~\ref{equation:cerenkov}, $S_e$ is in units of MeV$^2$, the frequency $\omega$ and $E_e$ are both in MeV, and the frequency integral is performed over the range 250--800 nm. The dispersion $n(\omega)$ comes from our ellipsometric measurements of the refractive indices of SNO\protect\raisebox{.15ex}{+} and EJ-301 LS above 210 nm \cite{refractive}. The four fit parameters were $A$, $B$, $C$ and $k_B$. \ref{section:lightyieldmodel} describes the relation between Eqs.~\ref{equation:lightyield} and \ref{equation:fitequation} in more detail. 
 
The fit results are shown in the upper plots of fig.~\ref{figure:resultsdatafit}. The scintillation and Cherenkov components are displayed as the dotted and dash-dotted curves, respectively. The solid-line curves show the best fits to the data. For the LAB-PPO sample, the fit returned $A=879\pm 12$, $B=0.78 \pm 0.05$, $C=-23.3\pm 2.2$, and $k_B\sim 74$ $\mu$m/MeV with a $\chi^2$/NDF of $12.3/22$. In the case of Nd-doped LAB-PPO, $A=797\pm 9$, $B=0.77\pm 0.03$, $C=-19.3\pm 1.7$, and $k_B\sim 91$ $\mu$m/MeV with a $\chi^2$/NDF of $25.7/25$. Assuming a scintillation light yield $L_s$ of $1\times 10^4$ photons/MeV, it appears that the Cherenkov light yield in the 250--800 nm window, $N_{C2}$, had to be scaled up by a factor of $L_sB/A \sim 9$ to explain our data. In comparison, we also fitted the data with:
\begin{equation}\label{equation:scintonly} {\cal L}(E_e)=A\cdot N_S^{\prime}(E_e)+C\end{equation}
which assumes a negligible Cherenkov component, and a scintillation light yield governed by Birks' law. The $\chi^2$/NDF for LAB-PPO and Nd-doped LAB-PPO were considerably worse: 264/23 and 492/26, respectively. 

For EJ-301, the results were $A=1558\pm 9$, $B=0.61\pm 0.04$, $C=-26.9\pm 3.5$, and  $k_B\sim 161$ $\mu$m/MeVwith a $\chi^2$/NDF of $14.4/14$. The $\chi^2$/NDF from the fit to Eq.~\ref{equation:scintonly} was $51.6/15$.

Therefore, as in the KamLAND case, we find that the non-linearity of the SNO\protect\raisebox{.15ex}{+} LS cannot be accounted for solely by Birks' law for ionization quenching. A combination of quenching and Cherenkov light excitation seems to better describe our measurements. For the SNO\protect\raisebox{.15ex}{+} LS, the dominant contribution to the non-linearity appears to come from the Cherenkov component. This causes an increase in light yield above the Cherenkov threshold ($\sim$0.2 MeV).
 
 \section{Conclusions}\label{section:conclusion}
 To summarize, we investigated the dependence of the light yield on electron energy for EJ-301 and two LAB-based scintillators with a Compton spectrometer apparatus, whose linear response was verified. The spectrometer allowed us to probe a large number of electron kinetic energy values between 0.09 and 3 MeV. We found that both LAB scintillator samples demonstrate an enhancement in the light yield in the vicinity of the Cherenkov threshold for electrons. This was also observed, albeit to a lesser extent, in the EJ-301 measurements. We provided a simple model to explain these observations, which deviate from the usual premise of LS electron energy scale linearity. 
 
 \section*{Acknowledgements}
 This work was supported by the United States Department of Energy grant \#DE-FG02-97ER41020. We thank our Brookhaven National Laboratory collaborators for sending the scintillator samples, M.~Chen and S.~Enomoto for helpful discussions, and the CENPA staff for technical assistance.
 
\appendix

\section{Cherenkov component}\label{section:lightyieldmodel}
Here we show that $N_C$ is directly proportional to $N_{C2}$, which is accurately calculable with Eq.~\ref{equation:cerenkov}. Consider the two wavelength ranges: (1) $\lambda\!<\!250$ nm, where Cherenkov yield cannot be fully calculated because the refractive index is unknown below 210nm, and (2) $\lambda\!>\!250$ nm. 
$N_C$ can be decomposed as:
 \begin{equation}\label{equation:cherenkov} N_{C}=f_{C2}\cdot N_{C2}+f_{C1}\cdot N_{C1}\end{equation}
where  $N_{C1}$ and $N_{C2}$ are the numbers of Cherenkov photons in regions 1 and 2, respectively. $f_{C1}$ and $f_{C2}$ are the fractions that are absorbed in regions 1 and 2, and re-emitted as one or more photons of longer, detectable, wavelengths. Since $f_{C1}$ and $f_{C2}$ have not been measured, and $N_{C1}$ is unknown, it is not possible to perform an absolute calculation of $N_C$.  However, assuming that the shape of the Cherenkov spectrum is independent of electron energy, $(f_{C1}N_{C1}/N_{C2}+f_{C2})$ is constant at all $E_e$. Thus, $N_C \propto N_{C2}$.

\end{document}